\documentclass{article}
\usepackage{spconf,amsmath,graphicx, flushend}
\usepackage{algorithm}
\usepackage{algorithmic}

\usepackage{amssymb}

\title{Adaptive Low Rank and Sparse Decomposition of Video using Compressive Sensing}
\name{Fei Yang$^1$  \hspace{3mm} Hong Jiang$^2$  \hspace{3mm} Zuowei
Shen$^3$ \hspace{3mm} Wei Deng$^4$ \hspace{3mm} Dimitris Metaxas$^1$
}

\address{$^1$Rutgers University  \hspace{3mm} $^2$Bell Labs \hspace{3mm} $^3$National University of Singapore\hspace{3mm}$^4$Rice University}
\begin{document}
%
\maketitle
\begin{abstract}
We address the problem of reconstructing and analyzing surveillance videos using compressive sensing. We develop a new method that performs video reconstruction by low rank and sparse decomposition adaptively. Background subtraction becomes part of the reconstruction. In our method, a background model is used in which the background is learned adaptively as the compressive measurements are processed.  The adaptive method has low latency, and is more robust than previous methods. We will present experimental results to demonstrate the advantages of the proposed method.
\end{abstract}
\vspace{-2mm}
\section{Introduction}
\vspace{-2mm}

In video surveillance, video signals are captured by cameras and transmitted to a processing center, where the videos are monitored and analyzed. Given a large number of cameras installed in public places, an enormous amount of data is generated and needs to be transmitted in the network, raising a high risk of network congestion. Therefore, it is highly desirable to compress the video signals transmitted in the network.

The recently introduced compressive sensing theory establishes that if a signal has a sparse representation in some basis, then it can be reconstructed from a small set of linear measurements \cite{Candes2006}\cite{Donoho06}. The number of measurements can be much smaller than that required by Nyquist sampling rate. Since videos are known to have a sparse representation in some transform basis (e.g. total variation, wavelet or framelet, etc.), the compressive sensing theory can be applied to compress video at the cameras, for example to acquire video by compressive measurements which can then be used to reconstruct the video \cite{jiang2012a}\cite{li2013}.

\begin{figure}[t]
\begin{center}
\includegraphics[width=1.0\linewidth]{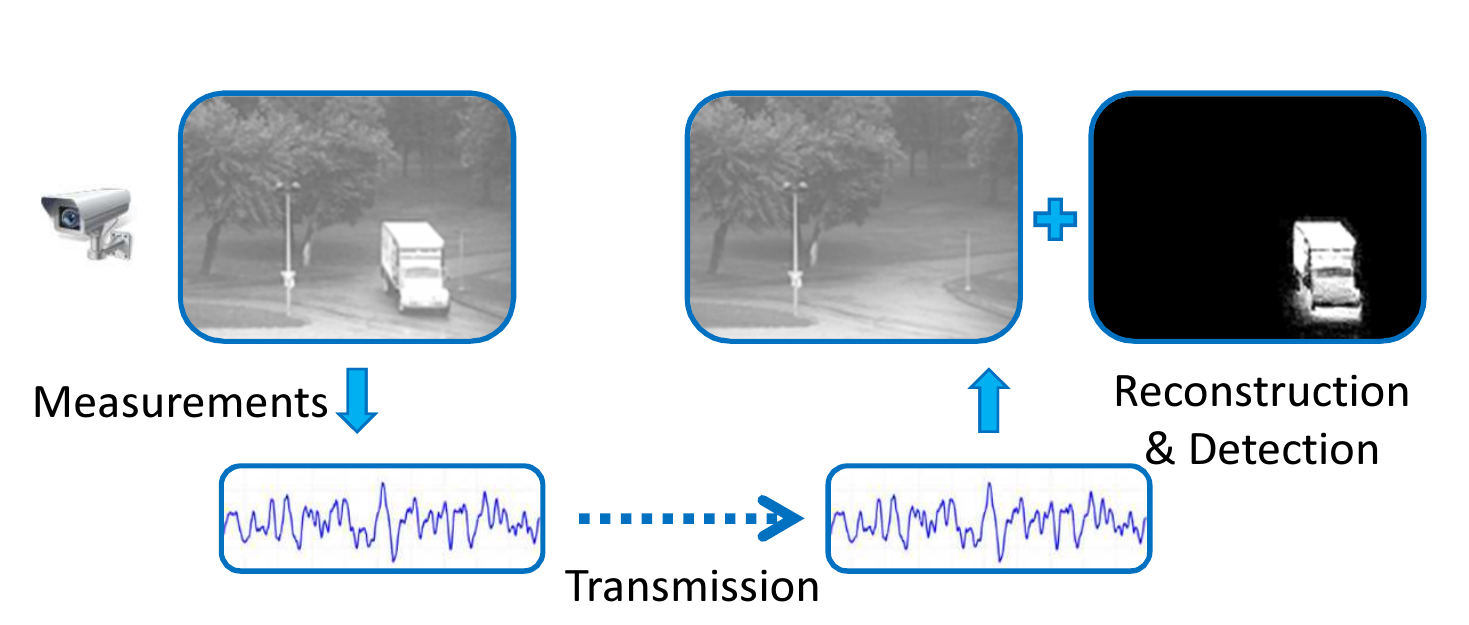}
\caption{The framework of the compressive sensing surveillance system. The video is compressed by using random projections, and then transmitted to the processing center. The frames are reconstructed and the moving objects are detected simultaneously.}
\end{center}
\label{fig:flowchar}
\vspace{-10mm}
\end{figure}

In this paper, we developed a framework for processing surveillance video using compressive measurements. Our system is shown in Fig.~\ref{fig:flowchar}. At the camera, the video captured by a surveillance camera is either acquired \cite{huang2013} as, or transformed to, the low dimensional measurements by using random projections. At the processing center, the frames of the video are reconstructed, and the moving objects are detected at the same time.

Our method is based on three observations: 1). The background is nearly static over a short period. Thus the background images lie in a low dimensional subspace. 2). Natural images are sparse in a transform, such as tight wavelet frame, domain. 3). Generally the moving objects only occupies a small portion of the field of view of a surveillance camera. Based on these observations, we use a low rank model for background and a sparse model for moving objects. The reconstruction of background and moving objects is performed by a low rank and sparse decomposition similar to \cite{candes09}\cite{Jiang2012}.

In the low rank model of \cite{Jiang2012}, a large number of frames of video must be used in order to properly reconstruct the background because the low rank and sparse decomposition computes background frames as a low rank basis of the space spanned by the incoming video frames. This results in a long latency in the reconstruction.

In this paper, we introduce an adaptive background model in which the low rank and sparse decomposition is performed with a small number of video frames. This significantly reduces latency. In this adaptive method, the video frames are reconstructed by a few frames at a time. In each reconstruction, the compressive measurements from a small number of video frames are used to perform the low rank and sparse decomposition which produces a set of background frames. The background frames are further processed and the results are used in the low rank and sparse decomposition for the next set of frames. Therefore, effectively, a large number of background frames are participated (although not explicitly used) in the computation of the low rank and sparse decomposition at each reconstruction, since the background frames from previous reconstructions are used. This makes it possible to accurately reconstruct background frames even with a small number of frames processed each time. The proposed method handles background changes very well because it is adaptive. Furthermore, the method reduces latency and computational complexity significantly.

In the remaining parts of the paper, we first introduce previous work related to our study. Then we introduce the framework of our video reconstruction method, followed by the background model and its adaption algorithm. The experimental results are given at the end.

\vspace{-2mm}
\section{Related work}
\vspace{-2mm}

\hspace{4.5mm}
\textbf{Background subtraction}.
There has been extensive study on background subtraction from original videos \cite{brutzer2011}. The earliest background subtraction methods use frame difference to detect foreground \cite{jain1979}. Subsequent approaches aimed to model the variations and uncertainty in background appearance, such as mixture of Gaussian \cite{stauffer2000} and non-parametric kernel density estimation \cite{elgammal2002}. Currently state-of-art background subtraction methods are able to get satisfactory results for stationary cameras. However, these methods cannot be applied to compressive measurements.

\textbf{Sparse reconstruction}.
Cevher et al. \cite{cevher2008} casted the background subtraction as a sparse approximation problem and solved it based on convex optimization. Their method relies on a background model trained from pure background frames, which requires the prior knowledge of the background. Jiang et al. \cite{Jiang2012} developed a low rank and sparse decomposition based approach to detect moving objects from a video. Their method solves all the frames at the same time, which results in a long latency and expensive computational cost. In contrast, the approach in this paper does not require a clean background for training, and it reconstructs background adaptively, with a small number of frames of video processed at a time. This  reduces latency and complexity.

\vspace{-2mm}
\section{Low rank and sparse decomposition}
\vspace{-2mm}
\subsection{Compressive measurements}
\vspace{-2mm}

We consider a video consisting of $m$ frames. Each frame has a total of $n$ pixels. Let $x_j \in \Re^n$ be a vector formed by concatenating all pixels in frame $j$. Let $X=[x_1,...,x_m] \in \Re^{n \times m}$ be a matrix containing $m$ columns representing the $m$ frames of the video. Let $\Phi \in \Re^{ r\times n}$ be a sensing matrix. The compressive measurements of $X$ are defined as
\begin{equation}
y=\Phi \circ X \triangleq [\Phi x_1,...,\Phi x_m],
\label{eq:measure}
\end{equation}
where $y\in \Re^{ r\times n}$ is a matrix of measurements, with a much smaller row dimension than $X$, i.e., $r\ll n$. Each column of $y$ contains $r$ measurements of a frame of video. In our work, $\Phi$ is composed of a set of $r$ randomly permutated rows of Walsh-Hadamard matrix.

\vspace{-2mm}
\subsection{Reconstruction}
\vspace{-2mm}

Given the measurements $y$ , we want to reconstruct the original video $X$. $X$ can be decomposed into background matrix $X_1$ and foreground matrix $X_2$:
\begin{equation}
X = X_1 + X_2 .
\label{eq:X1X2}
\end{equation}
In above, $X_1$ is a matrix each column of which is formed from the pixels of a background frame of the video. Similarly, $X_2$ is a matrix each column of which is formed from the pixels of a foreground frame of the video.
Thus the objective is to solve $X_1$ and $X_2$, satisfying Eqs.~(\ref{eq:measure}) and (\ref{eq:X1X2}). Apparently, this is an ill-posed problem which has infinite number of solutions. Therefore, we need some prior knowledge to find a proper solution.

\textbf{Low rank background.} We assume the background images have relative small changes over a short period, then the background matrix $X_1$ should have a low rank \cite{candes09}. We use the nuclear norm to measure the rank of this matrix, which is defined as the sum of single values $\sigma_i$:
\begin{equation}
||X_1||_* = trace(\sqrt{X_1 X_1^T}) = \sum_{i} \sigma_i.
\end{equation}

\textbf{Sparsity in transformed domain}. Previous work shows that natural images can be sparsely represented in a transformed space. We assume each background frame is sparse under transform $W_1$, and each foreground frame is sparse under a transform $W_2$ \cite{Jiang2012}. We use the the $l_1$-norm to measure the sparsity of the transformed background and foreground: $||W_1\circ X_1||_1$, $||W_2\circ X_2||_1$, where the $l_1$-norm is defined as
\begin{equation}
||Z||_1\triangleq \sum_{i} \sum_{j} |z_{ij}|, \; \; Z=[z_{ij}].
\label{eq:L1}
\end{equation}

\textbf{Sparse foreground}. We also assume the foreground only occupies a small portion of a frame, and therefore,  we can also use $l_1$-norm as defined in Eq.~(\ref{eq:L1}) to measure the its sparsity: $||X_2||_1$.

Given these prior assumptions, $X_1$ and $X_2$ can be reconstructed by solving the following optimization problem:
\begin{eqnarray}
(X_1, X_2) = \underset{X_1,X_2}{\arg\min} \hspace{2mm} \mu_1||X_1||_* + \mu_2||W_1\circ X_1||_1  \label{eq:energy1} \\
                                                     + \mu_3||W_2\circ X_2||_1 + \mu_4||X_2||_1  \nonumber
\label{eq:opt1}
\end{eqnarray}
\vspace{-6mm}
\begin{eqnarray}
\hspace{-10mm} \text{such that} \hspace{10mm}
y=\Phi \circ (X_1+X_2). \nonumber
\end{eqnarray}
In above,  $\mu_1$, $\mu_2$, $\mu_3$ and $\mu_4$ are nonnegative weights. $W_1$ and $W_2$ are sparsifying operators. In our system, we set $W_1=W_2=W$ as the framelet transform \cite{ron1997}\cite{Daubechies01}\cite{Jiang2012}.

Eq.~(\ref{eq:energy1}) is a convex problem, so standard convex optimization algorithms such as the interior point method \cite{bonnans2006} can be applied to find a solution. However, these standard methods are computationally expensive. Instead, as shown in \cite{cai2010singular}, singular value thresholding is more efficient for low rank decomposition. We apply the Augmented Lagrangian Alternating Direction (ALAD) algorithm introduced in Jiang et al.~\cite{Jiang2012}.

\vspace{-2mm}
\section{Adaptive reconstruction}
\vspace{-2mm}

To reconstruct the background and foreground by solving Eq.~(\ref{eq:energy1}), a large number of frames (i.e., the number of columns of $X_1$) are required. This is because the solution to Eq.~(\ref{eq:energy1}) captures the low rank basis in the space spanned by $X_1$. If the number of frames is small, a moving object may not change significantly, thus would be captured as part of background. Only when a large number of frames is used, the solution to (\ref{eq:energy1}) would reconstruct a background as expected. This is the reason that a large number of frames (i.e., $m>100$) must be used in \cite{Jiang2012}. The requirement for a large number of frames leads to a high latency in the reconstruction. In addition, the computational complexity of singular value thresholding is $O(m^3)$, which makes the algorithm highly computationally expensive as the $m$ becomes large.

In this section, we introduce an adaptive method to reduce both latency and complexity. In order to reduce latency, we want to process a small number of frames each time. However, in order to improve accuracy of reconstructed background, we still need a large number of columns to be present in the calculation of the nuclear norm $||\cdot||_*$. For this purpose, we augment $X_1$ by the previously calculated background frames. In other words, we replace $||X_1||_*$ in  Eq.~(\ref{eq:energy1}) by $||[M_b, X_1]||_*$ where $M_b$ is a matrix which is a model of previously calculated background frames, see equations (\ref{eq:mb}) and (\ref{eq:energy2}) below.

The key idea of the paper is that $M_b$, a representation of previously calculated background frames, is low dimensional and is computed adaptively as more frames are processed. $M_b$ may initially be an inaccurate approximation of the background frames, but as the adaptation proceeds, $M_b$ becomes progressively better representation of background frames. Furthermore, as background changes, $M_b$ changes accordingly with the background. Therefore, this method not only reduces latency and complexity, but also allows the reconstructed background frames to adapt quickly to the changes in the background of the video.

\vspace{-2mm}
\subsection{Augmented low rank decomposition}
\vspace{-2mm}

We assume that a set of $k$ background frames, $b_j$, are already computed in processing the previous frames. We put them in a background matrix defined as:
\begin{eqnarray}
X_b = [b_1, ..., b_k] \in \Re^{n \times k}.\nonumber
\end{eqnarray}

The augmented background matrix $\hat{X_1}$ is formed by combining the previously computed background matrix $X_b$ with the to-be-computed background $X_1$ of $m$ new frames:
\begin{eqnarray}
\hat{X_1} = [X_b, X_1] \in \Re^{n \times (k+m)}. \nonumber
\end{eqnarray}
The use of the augmented matrix makes it possible to reconstruct $X_1$, $X_2$ even if $X_1$ has a very small number of columns. We now require $\hat{X_1}$, instead of $X_1$,  to have a small rank. Therefore, the problem to solve is same as Eq.~(\ref{eq:energy1}) but with $||X_1||_*$ replaced by $||\hat{X_1}||_*$. By using $\hat{X_1}$, there is no need for $X_1$ to have a large number of columns. A theoretical justification is given in \cite{Jiang2013}.

\vspace{-2mm}
\subsection{Low dimensional background model}
\vspace{-2mm}

The computational complexity to optimize the low rank of $\hat{X}_1$ is $O(k+m)^3$, which grows quickly as frames are continuously being processed. Therefore, we need to find a lower dimensional background model $M_b\in \Re^{n \times p}$ from the computed background frames $X_b$, for a new augmented matrix:
$[M_b, X_1] \in \Re^{n \times (p+m)}$,
where $p \ll k$. We need to find $M_b$ such that the nuclear norm of $[M_b, X_1]$ could approximate the nuclear norm of $\hat{X}_1$, which leads to the following optimization problem:
\begin{eqnarray}
M_b = \underset{M_b}{\arg\min} \left | ||\hat{X}_1||_* - ||[M_b, X_1] ||_* \right |.
\end{eqnarray}
We perform SVD decomposition of the background matrix $X_b$, and form $M_b$ as
\begin{eqnarray}
X_b &=& U D V^T,  \label{eq:svdX} \\
 M_b &=& U_p D_p.
\label{eq:mb}
\end{eqnarray}
In Eqs.~(\ref{eq:svdX}) and (\ref{eq:mb}), $D$ is a diagonal matrix containing singular values of $X_b$, and $U$, $V$ are orthogonal matrices. $D_p$ is a diagonal matrix formed by the  $p$ largest single values, and $U_p$ is consist of the first $p$ columns of $U$.

Now, replacing $||X_1||_*$ by $||[M_b, X_1]||_*$ in Eq.~(\ref{eq:energy1}), we have the low latency reconstruction given as:
\begin{eqnarray}
(X_1, X_2) = \underset{X_1,X_2}{\arg\min} ~~\mu_1||[M_b, X_1]||_* + \mu_2||W_1\circ X_1||_1    \label{eq:energy2}\\
  \hspace{-10mm}                                      +  \mu_3||W_2\circ X_2||_1 + \mu_4||X_2||_1,         \nonumber
\end{eqnarray}
\vspace{-6mm}
\begin{eqnarray}
\hspace{-10mm} \text{such that}  \hspace{10mm}
y = \Phi \circ (X_1+X_2) .\nonumber
\end{eqnarray}

\vspace{-4mm}
\subsection{Optimization}
\vspace{-2mm}

We now use the Augmented Lagrangian Alternating Direction (ALAD) algorithm to solve the problem in Eq.~(\ref{eq:energy2}). The main difficulty is that the nuclear norm term involves an augmented matrix having both known columns and unknown columns. However, this can be handled by replacing the augmented matrix with a new variable. In addition, we introduce splitting variables to make the objective function separable. We perform variable substitution as below:
\begin{eqnarray}
Z_1 = [M_b, X_1],      ~      
Z_2 = W_1\circ X_1,     ~        
Z_3 = W_2\circ X_2.             
\end{eqnarray}
The ALAD optimization is shown in Alg.~\ref{fig:alg}. More details about the optimization framework can be found in \cite{Jiang2012}.

\begin{algorithm}
\begin{algorithmic}[h]
\STATE Initialize $Z_i^{(0)}, \Lambda_i^{(0)}$,
\REPEAT
    \vspace{1mm}
    \STATE Update $X_1$, $X_2$, while fixing $Z_i$ and $\Lambda_i$,
    \vspace{1mm}
    \STATE Update $Z_i$, while fixing $X_1$, $X_2$ and $\Lambda_i$,
    \vspace{1mm}
    \STATE Update $\Lambda_i$, while fixing $X_1$, $X_2$ and $Z_i$,
    \vspace{1mm}
\UNTIL converge
\end{algorithmic}
\caption{Reconstructing $X_1$ and $X_2$ using ALAD.} \label{fig:alg}
\end{algorithm}

\newcommand{\FigW}{0.63\linewidth}
\newcommand{\FigWW}{0.21\linewidth}
\begin{figure*}[t!]
\centering
\includegraphics[width=\FigWW]{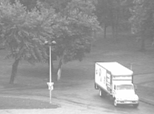}
\includegraphics[width=\FigWW]{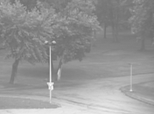}
\includegraphics[width=\FigWW]{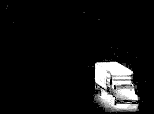}
\includegraphics[width=\FigWW]{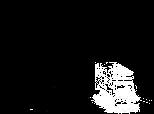}\\
\vspace{1mm}
\includegraphics[width=\FigWW]{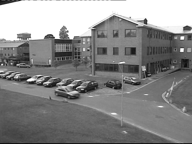}
\includegraphics[width=\FigWW]{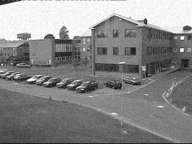}
\includegraphics[width=\FigWW]{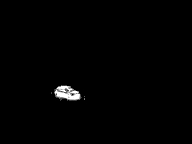}
\includegraphics[width=\FigWW]{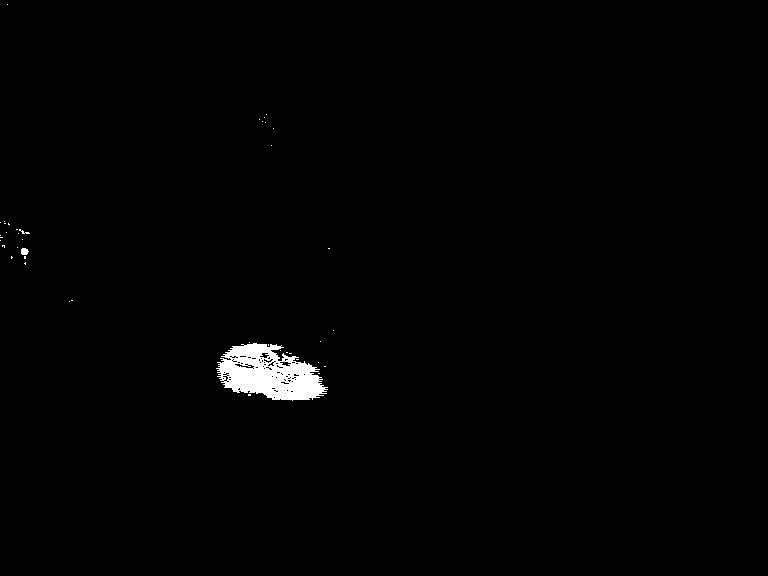}\\
\vspace{1mm}
\includegraphics[width=\FigWW]{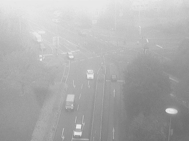}
\includegraphics[width=\FigWW]{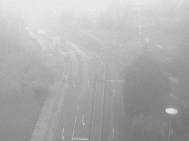}
\includegraphics[width=\FigWW]{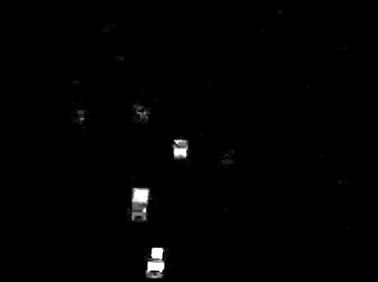}
\includegraphics[width=\FigWW]{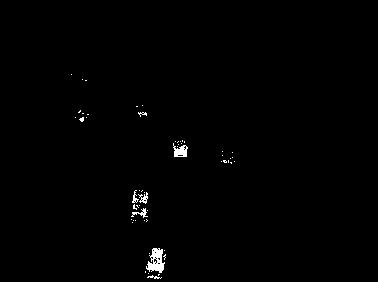}\\
\caption{Results of video reconstruction and background subtraction.
\textbf{Left}: original frames; \textbf{Middle}: background and
foreground reconstructed using the method of this paper; \textbf{Right}:
Foreground masks generated from original video with GMM.}
\label{fig:result} \label{fig:result}
\vspace{-2mm}
\end{figure*}

\vspace{-2mm}
\subsection{Updating the background model}
\vspace{-2mm}

With the previously computed $M_b$, Eq.~(\ref{eq:energy2}) can be used to compute current background frames $X_1$ by Alg.~\ref{fig:alg}. Then the question is, how do we update $M_b$ with current $X_1$ to obtain a new background model $M_b^{(new)}$ in order for us to solve Eq. (\ref{eq:energy2}) to reconstruct the next set of frames? We use an approach to update $M_b$ similar to the incremental SVD \cite{brand2006}. Given the SVD decomposition $X_b \approx U_p D_p V_p^T$, the decomposition of the augmented matrix with current background frames $X_1$ can be used to update $M_b$ as follows:
\begin{eqnarray}
\begin{bmatrix} U^{(new)} & D^{(new)}  \end{bmatrix}
        &=& svd(  \begin{bmatrix} w_b X_b                    & w_a X_1 \end{bmatrix}  ) ,          \nonumber \\
        &\approx& svd(  \begin{bmatrix} w_b U_p D_p V_p^T  & w_a X_1 \end{bmatrix}  ),  \nonumber \\
        &=& svd(  \begin{bmatrix} w_b U_p D_p            & w_a X_1 \end{bmatrix}  ),  \nonumber \\
        &=& svd(  \begin{bmatrix} w_b M_b                   & w_a X_1 \end{bmatrix}  ).  \nonumber \\
\vspace{2mm}
M_b^{(new)}&=&U^{(new)}_p  D^{(new)}_p. \label{eq:updateM}
\end{eqnarray}
In (\ref{eq:updateM}), $D^{(new)}_p$ is a diagonal matrix formed by the $p$ largest single values, and $U^{(new)}_p$ is consist of the first $p$ columns of $U^{(new)}$, similar to those in (\ref{eq:mb}). $w_a$ and $w_b$ are weights controlling the updating rate.

It is important to point out that in the update (\ref{eq:updateM}), the large matrix $V$ in SVD will never need to be computed, representing a significant reduction in complexity.

\vspace{-2mm}
\section{Experiments}
\vspace{-2mm}

We perform experiments on three video clips from PETS2001 database. The results are shown in Fig.~\ref{fig:result}. The first column shows the original frames. The second and third columns show backgrounds and foregrounds reconstructed by the method of this paper. We use 5\% measurements for the first two examples, and 10\% measurements in the last example which needs more measurements because a large number of small moving vehicles are difficult to detect from the background. Median filters are used to post-process the results of our method to reduce the noises. The last column shows the foregrounds generated by applying Gaussian Mixture model (GMM) \cite{stauffer2000}.

Fig.~\ref{fig:result} demonstrates that the results of our method are comparable to GMM. But our method are performed by only using 5\%-10\% of the original data, while GMM uses 100\%.

\vspace{-2mm}
\section{Conclusion}
\vspace{-2mm}

In this paper, we address the problem of reconstructing and analyzing surveillance videos from compressive measurements. We propose a method that simultaneously performs reconstruction and background subtraction with low latency. Our method is built on a background model, which is continuously updated as new frames are reconstructed. The experiments have proved the effectiveness and efficiency of the proposed method.

\bibliographystyle{IEEEbib}
\bibliography{adaptive}

\end{document}